\documentclass[journal]{IEEEtran}

\usepackage{amsmath,amsfonts,amssymb,amsthm,cite}

\def\R{\mathbb{R}}

\def\N{\mathbb{N}}

\def\e{\varepsilon}
\def\d{\delta}

\def\1{\mathds{1}}

\def\M{\mathcal{M}}
\def\P{\mathcal{P}}
\def\Q{\mathcal{Q}}

\def\B{\mathcal{B}}

\def\MA{M(X,\Sigma)}
\def\M1{M_1(X,\Sigma)}
\def\V{W}
\def\m{\mathfrak{m}}
\def\esssup{\mathrm{ess\text{ }sup}}

\def\H{\mathop{H}}
\def\h{\mathop{h}}
\def\hv{\mathop{h_W}}
\def\Hv{\mathop{H_W}}

\def\sh{\mathop{sh}}

\def\ldim{\underline{\mathrm{dim}}}
\def\udim{\overline{\mathrm{dim}}}
\def\lD{\underline{D}}
\def\uD{\overline{D}}
\def\dim{\mathrm{dim}}

\def\for{\mbox{  for }}

\newtheorem{observation}{Observation}[section]
\newtheorem{theorem}{Theorem}[section]
\newtheorem{proposition}{Proposition}[section]

\newtheorem{corollary}{Corollary}[section]

\theoremstyle{definition}
\newtheorem{definition}{Definition}[section]
\newtheorem{remark}{Remark}[section]
\newtheorem{example}{Example}[section]
\newtheorem{problem}{Problem}[section]

\begin{document}

\title{Entropy of the Mixture of Sources and Entropy Dimension}
%
%
%

\author{Marek~\'{S}mieja and Jacek~Tabor 
\thanks{The authors are with Institute of Computer Science, Jagiellonian University, Krak\'ow, Poland (e-mail: marek.smieja@ii.uj.edu.pl; jacek.tabor@ii.uj.edu.pl)}
}

\maketitle

\begin{abstract}
Suppose that we are given two sources $S_1$, $S_2$ and an ``error-control'' family $\Q$. We assume that we lossy-code $S_1$ with $\Q$-acceptable alphabet $\P_1$ and $S_2$ with $\Q$-acceptable alphabet $\P_2$.
Consider a new source $S$ which sends a signal produced by source $S_1$ with probability $a_1$  and by source $S_2$ with probability $a_2=1-a_1$. 
We provide a simple greedy algorithm which constructs a $\Q$-acceptable coding alphabet $\P$ of $S$ such that the entropy $\h(\P)$ satisfies:
$$
\h(\P) \leq a_1 \h(\P_1)+a_2\h(\P_2)+1.
$$
In the proof of the above formula the basic role is played by a new equivalent definition of entropy based on measures instead of partitions.

As a consequence we obtain an estimation of the entropy and R\'enyi entropy dimension of the convex combination of measures. In particular if probability measures $\mu_1,\mu_2$ have entropy dimension then
$$
\dim_E(a_1\mu_1+a_2\mu_2)=a_1\dim_E(\mu_1)+a_2\dim_E(\mu_2).
$$
In the case of probability measures in $\R^N$ this allows to link the upper local dimension at point with the upper entropy dimension of a measure by an improved version of Young estimation:
$$
\udim_E(\mu) \leq \int_{\R^N} \uD_{\mu}(x) d\mu(x),
$$
where $\uD_{\mu}(x)$ stands for upper local dimension of $\mu$ at point $x$.
\end{abstract}


%
\IEEEpeerreviewmaketitle

\section{Introduction}
%
%
%
%
\IEEEPARstart{T}{he} classical entropy introduced by C. E. Shannon \cite{Sh} and the entropy dimension\footnote{It is sometimes called R\'enyi information dimension.} defined by A. R\'enyi \cite{Re1} play a crucial role in information theory, coding, study of statistical and physical systems \cite{El, Go, Se, Wu}. In information theory, the entropy is understood as an absolute limit of the best possible lossless compression of any communication.  The entropy dimension in turn can be interpreted as a rate of convergence of the minimal amount of information needed to encode randomly chosen element with respect to maximal error decreasing to zero. 

\subsection{Motivation}

To explain our results, let us first recall that given a probability measure $\mu$ on a space $X$ and a countable partition $\P$ of $X$ into measurable sets,
we define the {\em entropy of $\mu$ with respect to $\P$} by the formula
\begin{equation}
\h(\mu;\P):=\sum_{P \in \P} \sh(\mu(P)),
\end{equation}
where $\sh(x):=-x\log_{2}x$. As we know the entropy corresponds to
the statistical amount of information given by optimal lossy-coding of $X$ by elements of partition $\P$, where $\P$ plays the role of 
the coding alphabet. Motivated by the idea of R\'enyi realized by the entropy dimension, we generalise the above formula for arbitrary 
measurable cover $\Q$ of $X$ by
\begin{equation}
\begin{array}{l}
\H(\mu;\Q):= \\[0.4ex]
\inf\{\h(\mu;\P) \, : \, \mbox{$\P$ is a partition of $X$ and $\P \prec \Q$}\}.
\end{array}
\end{equation}
The family $\Q$ is interpreted as a maximal error we are allowed to 
make in the lossy-coding. We accept only such coding alphabets 
$\P$, in which every element of $\P$ is a subset of a certain element of $\Q$
(if this is the case we say that $\P$ is $\Q$-acceptable).

\begin{remark}
The simplest natural case of such error-control family $\Q$ for classical random variables is given by the set $\B_{\d}$ of all intervals in $\R$ with length $\d$. Then to find $\H(\mu;\B_{\d})$ we need to consider the infimum of entropies  of all lossy-codings $\h(\mu;\P)$, where the elements of $\P$ have length not greater than $\d$. 

A. R\'enyi considered the above error-control family $\B_{\d}$ in his definition of entropy dimension \cite{Re1} (he also studied the more general case of 
metric spaces when $\B_{\d}$ denoted the family of all balls with radius $\d$).
One can also encounter in the general metric spaces the family of sets with diameter $\d$ or in the case of $\R^N$ of cubes with edge-length $\d$. 
\end{remark}

Our basic motivation in the paper was the following problem:
\begin{problem} \label{problemGlowny}
Suppose that we are given an error-control family $\Q$ and two sources $S_1$, $S_2$ in $X$ (represented by probability measures $\mu_1,\mu_2$ on $X$).
Let us consider a new source $S$ which sends a signal produced by source $S_1$ with probability $a_1$ 
and by source $S_2$ with probability $a_2=1-a_1$. Source $S$ is a mixture of $S_1$ and $S_2$. The question is what is the entropy of source $S$ with respect to the error $\Q$?

In other words we are interested in estimation of $\H(a_1\mu_1+a_2\mu_2;\Q)$ in terms of $\H(\mu_1;\Q)$ and $\H(\mu_2;\Q)$.
\end{problem}

\begin{observation}
Observe that if elements of $\Q$ are pairwise disjoint then the
answer to the above problem is trivial as by the subadditivity of the function $\sh$ we have
\begin{equation}
\H(\mu;\Q) = \h(\mu;\Q)=\sum_{Q \in \Q} \sh(\mu(Q))
\end{equation}
\begin{equation}
=\sum_{Q \in \Q}\sh(a_1\mu_1(Q)+a_2\mu_2(Q))
\end{equation}
\begin{equation}
\leq \sum_{Q \in \Q} \sh(a_1\mu_1(Q))+\sh(a_2\mu_2(Q))
\end{equation}
\begin{equation}
=a_1 \H(\mu_1;\Q)+a_2 \H(\mu_2;\Q)+\sh(a_1)+\sh(a_2).
\end{equation}

To see that the above estimation is sharp it is sufficient to consider a source
$S_1$ which sends only signal \verb#0# and source $S_2$ which sends signal \verb#1#. Clearly, $\H(S_1)=\H(S_2)=0$. Then the entropy of the source $S$ which sends signal generated by $S_1$ with probability
$a_1$ and $S_2$ with probability $a_2$ is exactly $a_1\H(S_1)+a_2\H(S_2)+\sh(a_1)+\sh(a_2)$.
\end{observation}

\subsection{Main Results}

In our main result, Theorem \ref{corEnt}, we show that the formula calculated in the above observation:
\begin{equation} \label{e0}
\begin{array}{l}
\H(a_1\mu_1+a_2\mu_2;\Q)  \\[0.4ex]
\leq a_1 \H(\mu_1;\Q)+a_2\H(\mu_2;\Q) + \sh(a_1)+\sh(a_2)
\end{array}
\end{equation}
is valid in the general case, that is when $\Q$ is an arbitrary measurable cover of $X$. The proof of our main result relies on a new definition of entropy based on measures instead of partitions, which we call {\em weighted entropy}. We provide an algorithm, which for given alphabets $\P_1, \P_2$ and measures $\mu_1, \mu_2$ allows to construct ``joint'' alphabet $\P$ satisfying above inequality.

\begin{remark}
We would like to add here that our idea of weighted entropy is indebted to the notion of weighted Hausdorff measures considered by J. Howroyd \cite{Ho1, Ho2}. The advantage of weighted Hausdorff measures over the classical ones is well-summarised by words of K. Falconer \cite[Introduction]{Ro}: "Recently, a completely different approach was introduced by Howroyd using weighted Hausdorff measures to enable the use of powerful techniques from functional analysis, such as the Hahn-Banach and Krein-Milman theorems." Making use of weighted Hausdorff measures Howroyd proves that
\begin{equation}
	\dim_H(X) + \dim_H(Y) \leq \dim_H(X \times Y),
\end{equation}
where $\dim_H(X)$ is the Hausdorff-Besicovitch dimension of $X$.
\end{remark}

For the precise definition of weighted entropy we refer the reader to the next section. We would only like to mention that, roughly speaking, weighted entropy provides the computation and interpretation of the entropy with respect to ``formal'' convex combination $a_1\P_1+a_2\P_2$, where $\P_1,\P_2$ are partitions (which clearly does not make
sense in the classical approach). This operation is crucial in the
proof of formula \eqref{e0}, whereas the second important part is played by Theorem \ref{wnWaz}, 
which proves that the weighted entropy is equal to the classical one.

As an easy consequence of \eqref{e0} in Theorem \ref{najwazniejszy} we obtain an estimation of the entropy dimension of the convex combination of measures. This result can be summarised as follows (see Corollary \ref{wnNajwaz}):\medskip
\begin{flushleft} \textit{Let $\mu_1$ and $\mu_2$ be probability measures which have entropy dimension and let $a_1, a_2 \in (0,1)$ be such that $a_1 + a_2 = 1$. Then $a_1 \mu_1 + a_2 \mu_2$ has entropy dimension and}
\begin{equation}
\dim_E(a_1\mu_1+a_2\mu_2)=a_1\dim_E(\mu_1)+a_2\dim_E(\mu_2),
\end{equation}
\textit{where $\dim_E(\cdot)$ stands for the entropy dimension of a given measure.}
\medskip
\end{flushleft}

In the case of measures in $\R^N$ this allows to combine the local upper dimension $\uD_\mu(\cdot)$ with the upper entropy dimension $\udim_E(\cdot)$ and improve Young estimation of the upper entropy dimension \cite{Fa}:
\begin{equation}
\udim_E(\mu) \leq \int_{\R^N} \uD_{\mu}(x) d\mu(x).
\end{equation}

\section{Weighted Entropy}

From now on, if not stated otherwise, we assume that $(X,\Sigma,\mu)$ is a probability space. The set of probability measures on $(X,\Sigma)$ will be denoted by $\M1$. When we consider a set of all measures then we will write $\MA$. 

\subsection{Shannon Entropy and Deterministic Coding}

We begin with the definition of $\mu$-partitions, which will play a role of a coding alphabet.
\begin{definition}
Let $\P \subset \Sigma$. We say that $\P$ is a \emph{$\mu$-partition (of $X$)} if $\P$ is countable family of disjoint sets and
\begin{equation}
\mu(X \setminus \bigcup_{P \in \P}P) = 0.
\end{equation}
\end{definition}
Consequently every element $x \in X$, which can be randomly drawn
(except for possibly elements of measure zero), is 
coded deterministically by the unique $P \in \P$ such that $x \in P$.

Then the entropy \cite{Sh} of $\mu$-partition is defined as follows:
\begin{definition}
Let $\P \subset \Sigma$ be a $\mu$-partition of $X$. We define \emph{$\mu$-entropy of $\P$} by
\begin{equation}
\h(\mu;\P):=\sum_{P \in \P} \sh(\mu(P)),
\end{equation}
where $\sh:[0,1] \to \R_+$ is the \emph{Shannon function}, i.e.
\begin{equation}
\sh(x):=\left\{
   \begin{array}{ll}
		- x \cdot \log_{2}(x) & \mbox{for } x \in (0,1],\\ 
		0 & \mbox{for } x = 0.
	\end{array}
\right.
\end{equation}
\end{definition}
Let us mention that $\sh$ is a continuous, concave and subadditive function.

Classical $\mu$-entropy is defined with use of disjoint sets, which is a very restrictive condition. It implies that we have fixed one alphabet $\P$ in our lossy-coding. However, this alphabet does not have to be optimal. In other words, there may exists another $\Q$-acceptable alphabet $\P'$, which provides less entropy than $\P$ (we assume that $\P$ is also $\Q$-acceptable). Thus it would be better to make a coding with use of $\P'$ rather than with $\P$. Therefore we will generalise the entropy for any error-control family. The error-control family can be an arbitrary family of measurable subsets of $X$. 

We say that family $\P$ is finer than $\Q$ (which we write $\P \prec \Q$) if for every $P \in \P$ there exists $Q \in \Q$ such that $P \subset Q$. When $\P$ is interpreted as a coding alphabet we may simply say that $\P$ is $\Q$-acceptable.

\begin{definition} \label{entDef1} 
Let $\Q \subset \Sigma$. We define the \emph{$\mu$-entropy of $\Q$} by
\begin{equation} \label{eq2}
\begin{array}{l}
\H(\mu;\Q):=\\[0.4ex]
\inf\{\h(\mu;\P) \in [0,\infty] \, : \, 
\mbox{$\P$ is a $\mu$-partition and $\P \prec \Q$}\}.
\end{array}
\end{equation}
\end{definition}

Observe that if there is no $\mu$-partition finer than $\Q$ then
directly from the definition\footnote{We put $\inf(\emptyset) = \infty$.} $\H(\mu;\Q)=\infty$. Moreover, if $\Q$ itself is a $\mu$-partition of $X$ then trivially\footnote{We can consider another $\mu$-partition $\P \prec \Q$ of $X$ but due to subadditivity of $\sh$ we get $\h(\mu;\Q) \leq \h(\mu;\P)$.} $\H(\mu;\Q)=\h(\mu;\Q)$. This observation implies that $\mu$-entropy $\H$ of $\Q$ is defined properly for $\mu$-partitions as well as for families of measurable subsets of $X$.

\subsection{Weighted Entropy and Random Coding}

Motivation of the weighted entropy is the following observation. Given error-control family $\Q$ in the classical approach we consider only $\Q$-acceptable deterministic codings $\P$. More precisely we always code a point $x \in X$ by the unique $P_x \in \P$ such that $x \in P_x$. 

However, if we do not insist on being deterministic in our coding, we could alternatively encode point $x$ by another set $P' \in \Sigma$ such that $x \in P'$ and for which there exists $Q' \in \Q: P' \subset Q'$. In this subsection we formalise this idea, namely we do not fix a $\Q$-acceptable alphabet $\P$ but we allow any random coding demanding only that $x$ can be encoded by $Q \in \Q$ iff $x \in Q$. Such a random coding might theoretically give lower entropy than the original one.

We make it precise in the following way. We define the space of functions from a family of measurable subsets of $X$ into a set of measures on $X$:
\begin{equation}
\begin{array}{l}
\V(\mu;\Q):=\{\m:\Q \ni Q \rightarrow \m_Q \in \MA : \\[0.4ex]
\m_Q(X \setminus Q) = 0 \text{ for every $Q \in \Q$ and } \sum_{Q \in \Q} \m_Q = \mu \}.
\end{array}
\end{equation}
Thus given $\m \in \V(\mu;\Q)$ and $Q \in \Q$, the value of $\m_Q(X)$ denotes the probability that an arbitrary point $x \in X$ is coded by $Q$ (and in that case $x \in Q$ with probability one). Observe also that every function $\m \in \V(\mu;\Q)$ is non-zero on at most countable sets of $\Q$.

Finally we define weighted $\mu$-entropy of a given $\m \in \V(\mu;\Q)$:
\begin{definition}
Let $\Q \subset \Sigma$. We define the {\em weighted $\mu$-entropy} of $\m \in \V(\mu;\Q)$ by
\begin{equation}
\hv(\mu; \m):=\sum_{Q \in \Q} \sh(\m_Q(X)) \text{.}
\end{equation}
The weighted $\mu$-entropy of $\Q$ is
\begin{equation}
\Hv(\mu; \Q) := \inf \{ \hv(\mu; \m) \in [0,\infty] : \m \in \V(\mu,\Q) \} \text{.}
\end{equation}
\end{definition}

The following remark explains the importance of the formulation of weighted entropy. 
\begin{remark}
Given functions $\m_1,\m_2 \in \V(\mu;\Q)$ and numbers $a_1,a_2 \in [0,1]$ such that $a_1 + a_2 = 1$ we are allowed to perform convex combinations $a_1 \m_1+ a_2 \m_2$ in the space $\V(\mu;\Q)$. Therefore we can compute the weighted $\mu$-entropy of a combination $\hv(\mu; a_1 \m_1 + a_2 \m_2)$ while the symbol $\h(\mu;a_1 \P_1 + a_2 \P_2)$ does not make sense for $\mu$-partitions $\P_1,\P_2$. This property will help us to find an estimation of entropy of convex combination of measures $\H(a_1 \mu_1 + a_2 \mu_2; \Q)$ for $\Q \subset \Sigma$.
\end{remark}


\subsection{Classical Entropy Equals Weighted}

In this section we show that the classical $\mu$-entropy of a family of measurable sets $\Q$ equals to the weighted $\mu$-entropy of $\Q$, i.e.
\begin{equation}
\Hv(\mu;\Q)=\H(\mu;\Q).
\end{equation}
It seems natural that every deterministic coding is a particular case of a random one. 
We will show it in the following proposition.

Let us denote the restriction of measure $\mu$ to $A \in \Sigma$ by 
\begin{equation}
\mu_{|A}(B):=\mu(A \cap B)
\end{equation}
for every $B \in \Sigma$.

\begin{proposition} \label{prop}
Random way of coding allows possibly more freedom than the deterministic one, i.e.
\begin{equation}
\Hv(\mu; \Q) \leq \H(\mu; \Q)
\end{equation}
for every family $\Q \subset \Sigma$.
\end{proposition}

\begin{IEEEproof}
Let us first observe that if there is no $\mu$-partition finer than $\Q$ then $\H(\mu;\Q)=\infty$ and the inequality holds trivially.

Thus let $\P$ be a $\mu$-partition finer than $\Q$. As $\P \prec \Q$, for every $P \in \P$ there exists $Q \in \Q$ such that $P \subset Q$. Hence we obtain a mapping $\pi: \P \to \Q$ satisfying $P \subset \pi(P)$. We define the family 
\begin{equation}
\P_\Q := \{P_Q\}_{Q \in \Q},
\end{equation}
where $P_Q:=\bigcup\limits_{P:\pi(P)=Q}P$. Let us notice that $\P_\Q$ is a $\mu$-partition and $\P \prec \P_\Q \prec \Q$. Finally, we put $\m:\Q \ni Q \rightarrow \mu_{|P_Q} \in \MA$.

Since $\P_\Q$ is a $\mu$-partition and $P_Q \subset Q$ for every $Q \in \Q$ then
\begin{equation}
\sum_{Q \in \Q} \m_Q(X) = \sum_{Q \in \Q} \mu_{|P_Q}(Q) 
= \sum_{Q \in \Q} \mu(P_Q) = \mu(X).
\end{equation}
Moreover, for every $Q \in \Q$
\begin{equation}
\m_Q(X \setminus Q) = \mu_{|P_Q}(X \setminus Q) 
\leq \mu_{|Q}(X \setminus Q)= 0.
\end{equation}
Thus $\m \in \V(\mu;\Q)$. Making use of subadditivity of $\sh$ we obtain
\begin{equation}
\hv(\mu;\m) = \sum_{Q \in \Q}\sh(\m_Q(X)) 
= \sum_{Q \in \Q}\sh(\mu_{|P_Q}(X)) 
\end{equation}
\begin{equation}
= \sum_{Q \in \Q}\sh(\mu(P_Q)) 
= \sum_{Q \in \Q}\sh(\mu(\bigcup_{P: \pi(P)=Q}P)) 
\end{equation}
\begin{equation}
\leq \sum_{Q \in \Q} \sum_{P: \pi(P)=Q}\sh(\mu(P)) 
= \sum_{P \in \P} \sh(\mu(P)) = \h(\mu;\P).
\end{equation}
We conclude that $\Hv(\mu; \Q) \leq \H(\mu; \Q)$.
\end{IEEEproof}

The opposite inequality is more difficult to prove. To do this we will need an additional proposition. Given $\m \in \V(\mu;\Q)$ we will construct a $\mu$-partition $\P$ finer than $\Q$ with not greater entropy.

\begin{proposition} \label{waz}
Let $\Q =\{Q_i\}_{i \in I}$ be a family of measurable subsets of $X$, where either $I=\N$ or $I=\{1,\ldots,N\}$ for a certain $N \in \N$. Let $\m \in \V(\mu;\Q)$. We assume that
\begin{itemize}
 \item $\mu(X \setminus \bigcup\limits_{i \in I}Q_i) = 0$,
 \item the sequence $I \ni i \rightarrow \m_{Q_i}(X)$ is nonincreasing.
\end{itemize}
We define the family $\P=\{P_i\}_{i \in I} \subset \Sigma$ by the formula
\begin{equation}
P_1:=Q_1, \, P_i:=Q_i \setminus \bigcup_{k=1}^{i-1} Q_{k} \for i \in I, i \geq 2.
\end{equation}
Then $\P$ is a $\mu$-partition, $\P \prec \Q$ and
\begin{equation} \label{cosik}
\hv(\mu; \m) \geq \h(\mu;\P).
\end{equation}
\end{proposition}

\begin{IEEEproof}
Let us observe that by the definition of $\P$, we have $\P \prec \Q$. Moreover, since $\mu(X \setminus \bigcup\limits_{i \in I}Q_i)=0$ and $\bigcup\limits_{i \in I} P_i = \bigcup\limits_{i \in I} Q_i$, we get that $\P$ is a $\mu$-partition.

To prove (\ref{cosik}) we define sequences $(x_i)_{i \in I} \subset [0,1]$ and $(y_i)_{i \in I} \subset [0,1]$ by the formulas
\begin{equation}
x_i:=\m_{Q_i}(X)=\m_{Q_i}(Q_i), \, 
\end{equation}
\begin{equation}
y_i:=\mu(P_i)
\end{equation}
for $i \in I$. Then
\begin{equation}
\sum_{i \in I} x_i=\mu(X)=\sum_{i \in I} y_i.
\end{equation}
Directly from the assumption we conclude that $(x_i)_{i \in I}$ is a nonincreasing sequence. Moreover, for every $n \in I$:
\begin{equation}
\sum_{i = 1}^n x_i=\sum_{i = 1}^n \m_{Q_i}(Q_i) 
=(\sum_{i = 1}^n \m_{Q_i})(Q_1 \cup \ldots \cup Q_n)
\end{equation}
\begin{equation}
\leq \mu(Q_1 \cup \ldots \cup Q_n)
=\sum_{i = 1}^n \mu(P_i)=\sum_{i = 1}^n y_i.
\end{equation}
We have obtained that
\begin{equation} \label{wowo}
\sum_{i = 1}^n x_i \leq \sum_{i = 1}^n y_i \text{ for } n \in I.
\end{equation}
By applying the version of Hardy-Polya-Littlewood Theorem (see Appendix A for details) for sequences $(x_i)_{i \in I}$, $(y_i)_{i \in I}$ and the concave function $\sh$ we conclude that
\begin{equation}
\hv(\mu; \m)= \sum_{i \in I} \sh(\m_{Q_i}(X)) 
= \sum_{i \in I}\sh(x_i) 
\end{equation}
\begin{equation}
\geq \sum_{i \in I}\sh(y_i) 
= \sum_{i \in I}\sh(\mu(P_i))=\h(\mu;\P).
\end{equation}
\end{IEEEproof}

As a direct corollary we obtain that both random and deterministic coding provide the same entropy.
\begin{theorem} \label{wnWaz}
Let $\Q \subset \Sigma$. Then weighted entropy coincides with the classical entropy, i.e.
\begin{equation}
\Hv(\mu;\Q) = \H(\mu;\Q).
\end{equation}
\end{theorem}

\begin{IEEEproof}
Clearly by Proposition \ref{prop}, we get $\Hv(\mu;\Q) \leq \H(\mu;\Q)$.

To obtain the opposite inequality, let us first observe that if $\V(\mu;\Q) = \emptyset$ then $\Hv(\mu;\Q) = \infty$ and trivially $\Hv(\mu;\Q) \geq \H(\mu;\Q)$. 

We discuss the case when $\V(\mu;\Q) \neq \emptyset$. Let $\m \in \V(\mu;\Q)$ be arbitrary. We define the family of measurable subsets of $X$ by
\begin{equation}
\tilde{\Q} := \{Q \in \Q : \m_Q(X) > 0\}.
\end{equation}
Let us notice that $\tilde{\Q}$ is a countable family since $\sum\limits_{Q \in \tilde{\Q}} \m_Q(X) = 1$. Clearly, $\tilde{\m}:=\m_{|\tilde{\Q}} \in \V(\mu; \tilde{\Q})$. Moreover, $\tilde{\Q} \prec \Q$ and $\hv(\mu;\tilde{\m})=\hv(\mu;\m)$.

As $\tilde{\Q}$ is countable, we may find a set of indices $I \subset \N$ such that $\tilde{\Q}=\{Q_i\}_{i \in I}$ and the sequence $I \ni i \rightarrow \m_{Q_i}(X)$ is nonincreasing. Making use of Proposition \ref{waz} we construct a $\mu$-partition $\P \prec \tilde{\Q}$, which satisfies
\begin{equation}
\hv(\mu;\tilde{\m}) \geq \h(\mu;\P)
\end{equation}
This completes the proof since $\P \prec \tilde{\Q} \prec \Q$ and $\hv(\mu;\m) = \hv(\mu;\tilde{\m}) \geq \h(\mu;\P)$.
\end{IEEEproof}

As we proved the equality between classical and weighted entropy, we will use one notation $\H(\mu;\Q)$ to denote both classical and weighted $\mu$-entropy of $\Q \subset \Sigma$.


\section{Entropy of the Mixture of Sources}

\subsection{Estimation of the Entropy}

We return to Problem \ref{problemGlowny}. We are given two sources $S_1, S_2$, which are represented by probability measures $\mu_1, \mu_2$ respectively. Suppose that we have fixed error-control family $\Q \subset \Sigma$, which  defines the precision in the lossy-coding elements of $X$. Let us consider a new source $S$ which sends a signal produced by $S_1$ with probability $a_1$ and produced  by $S_2$ with probability $a_2$. We are interested in estimation of the entropy of $S$ (mixture of $S_1$ and $S_2$) with respect to $\Q$ in terms of $\H(\mu_1;\Q)$ and $\H(\mu_2;\Q)$. In other words we would like to measure how much memory we need to reserve for information from source $S$ providing that we know the mean amount of information needed to encode elements sent by $S_1$ and $S_2$ separately.

We will consider a general case: we assume $n \in \N$ sources $S_1,\ldots,S_n$. Let us begin with a proposition.
\begin{proposition} \label{propozycja}
Let $n \in \N$ and let $a_k \in (0,1)$ for $k \in \{1,\ldots,n\}$ be such that $\sum\limits_{k=1}^n a_k = 1$. Let $\{\mu_k\}_{k=1}^n \subset \M1$. We put $\mu := \sum\limits_{k=1}^n a_k \mu_k \in \M1$.
\begin{itemize}
\item
If $\P$ is a $\mu$-partition of $X$ then $\P$ is a $\mu_k$-partition of $X$ for $k \in \{1,\ldots,n\}$ and
\begin{equation} \label{pierwsza}
\h(\mu; \P) \geq \sum_{k=1}^n a_k \h(\mu_k; \P).
\end{equation}
\item If $\Q \subset \Sigma$ and $\m^k \in \V(\mu_k;\Q)$ for $k \in \{1,\ldots,n\}$ then $\m := \sum\limits_{k=1}^n a_k \m^k \in \V(\mu;\Q)$ and
\begin{equation} \label{druga}
\hv(\mu; \m) \leq \sum_{k=1}^n a_k \hv(\mu_k; \m^k)+\sum_{k=1}^n \sh(a_k).
\end{equation}
\end{itemize}
\end{proposition}

\begin{IEEEproof}
Clearly, $\P$ is a $\mu_k$-partition for every $k \in \{1,\ldots,n\}$. As a direct consequence of the concavity of the Shannon function we obtain that
\begin{equation}
\h(\mu;\P)=\sum_{P \in \P}\sh(\mu(P))
=\sum_{P \in \P}\sh(\sum_{k=1}^n a_k\mu_k(P))
\end{equation}
\begin{equation}
\geq \sum_{P \in \P} \sum_{k=1}^n a_k \sh(\mu_k(P)) 
= \sum_{k=1}^n a_k \h(\mu_k;\P)
\end{equation}
which proves (\ref{pierwsza}).

It is easy verify that $\m \in \V(\mu;\Q)$. To prove (\ref{druga}) we use subadditivity of the Shannon function and property: $\sh(ax)=a\sh(x) + x \sh(a)$.
\begin{equation}
\hv(\mu; \m) = \sum_{Q \in \Q} \sh(\sum_{k=1}^n a_k \m^k_{Q}(X)) 
\end{equation}
\begin{equation}
\leq \sum_{Q \in \Q} \sum_{k=1}^n \sh(a_k \m^k_{Q}(X))
\end{equation}
\begin{equation}
= \sum_{k=1}^n \sum_{Q \in \Q} [ a_k \sh(\m^k_{Q}(X)) + \sh(a_k) \m^k_{Q}(X) ] 
\end{equation}
\begin{equation}
= \sum_{k=1}^n a_k \hv(\mu_k; \m^k) + \sum_{k=1}^n \sh(a_k).
\end{equation}
\end{IEEEproof}

Making use of Proposition \ref{propozycja} we can estimate the entropy of convex combination of measures, which is the main result of the paper:
\begin{theorem} \label{corEnt}
Let $n \in \N$ and let $a_k \in [0,1]$ for $k \in \{1,\ldots,n\}$ be such that $\sum\limits_{k=1}^n a_k = 1$. Let $\{\mu_k\}_{k=1}^n \subset \M1$. If $\Q \subset \Sigma$ then
\begin{equation} \label{nierowWaz}
\H(\sum_{k=1}^n a_k \mu_k;\Q) \geq \sum_{k=1}^n a_k \H(\mu_k;\Q) 
\end{equation}
and
\begin{equation} \label{nierowWaz2}
\H(\sum_{k=1}^n a_k \mu_k;\Q) \leq \sum_{k=1}^n a_k \H(\mu_k;\Q) + \sum_{k=1}^n \sh(a_k).
\end{equation}
\end{theorem}

\begin{IEEEproof}
We consider the case when all considered entropies are finite because if $\H(\mu_k;\Q) = \infty$ for a certain $k\in\{1,\ldots,n\}$ then also $\H(\mu;\Q) = \infty$ and the proof is completed. Moreover, without loss of generality, we may assume that $a_k \neq 0$ for every $k \in \{1,\ldots,n\}$.

We denote $\mu := \sum\limits_{k=1}^n a_k \mu_k$. Let $\e>0$ be arbitrary. By the definition of entropy, we find a $\mu$-partition $\P$ finer than $\Q$ such that 
\begin{equation} \label{gwiazdka}
\H(\mu;\Q) \geq \h(\mu;\P)-\e. 
\end{equation}
Then by Proposition \ref{propozycja}, we have
\begin{equation}
\h(\mu;\P) = \h(\sum_{k=1}^n a_k \mu_k; \P) 
\end{equation}
\begin{equation}
\geq \sum_{k=1}^n a_k \h(\mu_k;\P) \geq \sum_{k=1}^n a_k \H(\mu_k;\Q).
\end{equation}
Consequently by (\ref{gwiazdka}),
\begin{equation}
\H(\mu;\Q) \geq \h(\mu; \P) -\e 
\geq \sum_{k=1}^n a_k \H(\mu_k;\Q) - \e.
\end{equation}

We prove the second inequality. Again by the definition, for each $k \in \{1,\ldots n\}$ we find $\m^k \in \V(\mu_k;\Q)$ such that
\begin{equation} \label{2gwiazdki}
\hv(\mu_k; \m^k) \leq \H(\mu_k;\Q) + \frac{\e}{n}.
\end{equation}
Then by Proposition \ref{propozycja} and (\ref{2gwiazdki}), we obtain 
\begin{equation}
\H(\mu;\Q) \leq \hv(\mu; \sum_{k=1}^n a_k \m^k) 
\end{equation}
\begin{equation}
\leq \sum_{k=1}^n [a_k \hv(\mu_k; \m^k) + \sh(a_k)] 
\end{equation}
\begin{equation}
\leq \sum_{k=1}^n [a_k \H(\mu_k;\Q) + \sh(a_k)] + \e ,
\end{equation}
which completes the proof as $\e>0$ was an arbitrary number.
\end{IEEEproof}
Clearly, $\sum\limits_{k=1}^n \sh(a_k) \leq \log_2(n)$. Thus the assertion (\ref{nierowWaz2}) of Theorem \ref{corEnt} can be also rewritten as 
\begin{equation}
\H(\sum_{k=1}^n a_k \mu_k;\Q) \leq \sum_{k=1}^n a_k \H(\mu_k;\Q) + \log_2(n).
\end{equation}
When we consider a combination of two probability measures then we get directly:
\begin{corollary}
Let $a_1, a_2 \in (0,1)$ be such that $a_1 + a_2 = 1$. Given probability measures $\mu_1, \mu_2$ and a family of measurable subsets $\Q$ of $X$, we have
\begin{equation}
\H(a_1 \mu_1 + a_2 \mu_2;\Q) \geq a_1 \H(\mu_1;\Q) + a_2 \H(\mu_2;\Q),  
\end{equation}
\begin{equation} \label{algor}
\H(a_1 \mu_1 + a_2 \mu_2;\Q) \leq a_1 \H(\mu_1;\Q) + a_2 \H(\mu_2;\Q) + 1.
\end{equation}
\end{corollary}


\subsection{Practical Algorithm for Finding ``Joint'' Coding Alphabet of the Mixture of Sources}

A practical question is how to construct $\Q$-acceptable coding alphabet $\P$ form given alphabets $\P_1$ and $\P_2$ such that
\begin{equation}
\begin{array}{l}
\h(a_1 \mu_1 + a_2 \mu_2;\P) \leq \\[0.4ex]
\leq a_1 \h(\mu_1;\P_1) + a_2 \h(\mu_2;\P_2) 
+ \sh(a_1) + \sh(a_2).
\end{array}
\end{equation} 
For the case of simplicity we consider only the case when $\P_1$ and $\P_2$ are finite families. 

Based on Propositions \ref{waz} and \ref{propozycja} it is not difficult to deduce the following simple, but general, greedy algorithm for constructing such an alphabet $\P$.

ALGORITHM:
\begin{enumerate}
\item $i=0$; \\
		$\P^0=\P_1 \cup \P_2$;
\item IF $\P^i$ is empty GOTO STEP 4; \\
		ELSE find a set $\bar P_i \in \P^i$ which maximises the value of
$$
\P^i \ni P \to a_1\mu_1(P)+a_2\mu_2(P);
$$ 
IF maximum equals zero GOTO STEP 4;
\item $\P^{i+1}=\{P \setminus \bar P_i \, : \, P \in \P^i\}$;\\
	 	$i=i+1$; \\
		GOTO STEP 2;
\item $\P=\{\bar P_0,\bar P_1,\ldots,\bar P_{i-1}\}$; 
		\\END.
\end{enumerate}
Clearly, this algorithm can be directly adopted for more than two sources in $X$.

Let us look how the above algorithm works in practice.
\begin{example}
Let $X=[0,2]$. We consider two measures $\mu_1:[0,1] \rightarrow \R$ and $\mu_2:[\frac{1}{10}, \frac{11}{10}] \rightarrow \R$ given by
\begin{equation}
\mu_1(A) = \int_A 1 \, dx \text{, } \mu_2(A) = 2 \int_A (x - \frac{1}{10})\, dx.
\end{equation}
As an error-control family $\Q$ we take the family of all intervals contained in $[0,2]$ with length not greater than $\frac{2}{5}$. We consider coding alphabets:
\begin{equation}
\begin{array}{l}
\P_1=\{[0,\frac{2}{5}),[\frac{2}{5},\frac{3}{5}),[\frac{3}{5},\frac{4}{5}),[\frac{4}{5},1]\}, \\[0.4ex]
\P_2=\{[\frac{1}{10},\frac{1}{2}),[\frac{1}{2},\frac{7}{10}),[\frac{7}{10},\frac{9}{10}),[\frac{9}{10},\frac{11}{10}]\}.
\end{array}
\end{equation}
Mixture of sources is given by probabilities $a_1=2/5$ and $a_2=3/5$. 

The algorithm presented above produces following $\Q$-acceptable alphabet of the mixture: 
\begin{equation}
\begin{array}{l}
\P=\{[0,\frac{1}{10}),[\frac{1}{10},\frac{1}{2}),[\frac{1}{2},\frac{3}{5}),[\frac{3}{5},\frac{4}{5}),[\frac{4}{5},1],(1,\frac{11}{10}]\}.
\end{array}
\end{equation}
We get the entropies: 
\begin{equation}
a_1 \h(\mu_1;\P_1) + a_2 \h(\mu_2;\P_2) \approx 1.93,
\end{equation}
\begin{equation}
a_1 \h(\mu_1;\P_1) + a_2 \h(\mu_2;\P_2) + \sh(a_1) + \sh(a_2) \approx 2.9,
\end{equation}
\begin{equation}
\h(a_1 \mu_1 + a_2 \mu_2; \P) \approx 2.36.
\end{equation}
\end{example}

As we see, we have obtained a reasonable coding method for finding joint alphabet of the mixture of sources.


\section{R\'enyi Entropy Dimension} \label{sekcjaDim}

From now on we always assume that $X$ is a metric space and $\Sigma$ contains all Borel subsets of $X$.

\subsection{Entropy Dimension of Convex Combination of Measures}

Entropy of a probability measure $\mu$ with respect to the error-control family $\Q \in \Sigma$ identifies minimal amount of information needed to encode an arbitrary element of $X$ with error $\Q$. R\'enyi entropy dimension in turn gives the rate of convergence of this quantity when error is decreasing. Thus it is also important to estimate the entropy dimension of convex combination of measures. Making use of Theorem \ref{corEnt} it is quite simple.

Given $\d>0$ let us denote a family of all balls in $X$ with radius $\d$ by
\begin{equation}
\B_\d := \{B(x,\d): x \in X\},
\end{equation}
where $B(x,\d)$ is a closed ball centred at $x$ with radius $\d$.

We consider $\B_\d$ as an error-control family. If we want to code a point $x \in X$ by a certain ball $B(q,\d)$ then we may code it in fact by its centre $q$. Thus the error we make, simply equals to the distance between $x$ and $q$. Consequently, the family $\B_\d$ allows to code points from $X$ with error not greater than $\d$. 

For the convenience of the reader let us recall the definition of the entropy dimension \cite{Re1}. 
\begin{definition}
The \emph{upper and lower entropy dimension} of measure $\mu \in \M1$ are defined by
\begin{equation}
\udim_E(\mu):=\limsup_{\d \to 0} \frac{\H(\mu;\B_\d)}{-\log_{2}(\d)}, \,
\end{equation}
\begin{equation}
\ldim_E(\mu):=\liminf_{\d \to 0} \frac{\H(\mu;\B_\d)}{-\log_{2}(\d)}.
\end{equation}
If the above are equal we say that $\mu$ has the \emph{entropy dimension} and denote it by $\dim_E(\mu)$.
\end{definition}

We apply Theorem \ref{corEnt} for estimation of R\'enyi entropy dimension of convex combination of measures.
\begin{theorem} \label{najwazniejszy}
Let $n \in \N$ and let $a_k \in [0,1]$ for $k \in \{1,\ldots,n\}$ be such that $\sum\limits_{k=1}^n a_k = 1$. If $\{\mu_k\}_{k = 1}^n \subset \M1$ then
\begin{equation}
\ldim_E(\sum_{k=1}^n a_k \mu_k) \geq \sum_{k=1}^n a_k \ldim_E(\mu_k),
\end{equation}
\begin{equation}
\udim_E(\sum_{k=1}^n a_k \mu_k) \leq \sum_{k=1}^n a_k \udim_E(\mu_k).
\end{equation}
\end{theorem}
\begin{IEEEproof}
Let $\d \in (0,1)$ be given. By Theorem \ref{corEnt}, we have
\begin{equation}
\H(\sum_{k=1}^n a_k \mu_k; \d) \geq \sum_{k=1}^n a_k \H(\mu_k; \d) 
\end{equation}
and
\begin{equation}
\H(\sum_{k=1}^n a_k \mu_k; \d) \leq \sum_{k=1}^n a_k \H(\mu_k; \d) + \sum_{k=1}^n \sh(a_k).
\end{equation}
Dividing by $-\log_{2}(\d)$ and taking respective limits as $\d \to 0$, we obtain assertion of the theorem.
\end{IEEEproof}

\begin{corollary} \label{wnNajwaz}
Let $n \in \N$ and let $a_k \in [0,1]$ for $k \in \{1,\ldots,n\}$ be such that $\sum\limits_{k=1}^n a_k = 1$. Let $\{\mu_k\}_{k = 1}^n \subset \M1$. If every $\mu_k$ has entropy dimension for $k \in \{1,\ldots,n\}$ then $\sum\limits_{k=1}^n a_k \mu_k$ also has entropy dimension and 
\begin{equation}
\dim_E(\sum_{k=1}^n a_k \mu_k) = \sum_{k=1}^n a_k \dim_E(\mu_k).
\end{equation}
\end{corollary}

We generalise Theorem \ref{najwazniejszy} for the case of countable families of measures under an additional assumption that the upper box dimension of $X$ is finite. It will allow us to prove stronger version (see Corollary \ref{corRen}) of theorem proved by A. R\'enyi \cite[page 196]{Re1} concerning the entropy dimension of discrete measure. It is worth mentioning first the definition of upper box dimension \cite{Fal}. 

The upper box dimension of any non-empty bounded subset $S$ of $X$ is defined by
\begin{equation}
\udim_B(S):=\limsup_{\d \to 0} \frac{\log{N_\d(S)}}{-\log{\d}}, \,
\end{equation}
where $N_\d(S)$ denotes the smallest number of closed balls of radius $\d$ that cover $S$.

\begin{theorem} \label{countable}
We assume that $\udim_B(X) < \infty$. Let $\{\mu_k\}_{k=1}^\infty \subset \M1$ and let a sequence $(a_k)_{k=1}^\infty \subset [0,1]$ be such that $\sum\limits_{k=1}^\infty a_k = 1$. Then
\begin{equation}
\ldim_E(\sum_{k=1}^\infty a_k \mu_k) \geq \sum_{k=1}^\infty a_k \ldim_E(\mu_k) 
\end{equation}
and
\begin{equation}
\udim_E(\sum_{k=1}^\infty a_k \mu_k) \leq \sum_{k=1}^\infty a_k \udim_E(\mu_k).
\end{equation}
\end{theorem}

\begin{IEEEproof}
To prove first inequality we use Theorem \ref{najwazniejszy}. For every $N \in \N$ we have:
\begin{equation}
\ldim_E(\sum_{k=1}^\infty a_k \mu_k)
= \ldim_E( (\sum_{i=1}^N a_i) \sum_{k=1}^N \frac{a_k} {\sum_{j=1}^N a_j} \mu_k 
\end{equation}
$$
+ (\sum_{i=N+1}^\infty a_i) \sum_{k=N+1}^\infty \frac{a_k}{\sum_{j=N+1}^\infty a_j} \mu_k )
$$
\begin{equation}
\geq (\sum_{i=1}^N a_i)\ldim_E(\sum_{k=1}^N \frac{a_k}{\sum_{j=1}^N a_j} \mu_k) 
\end{equation}
$$
+ (\sum_{i=N+1}^\infty a_i)\ldim_E(\sum_{k=N+1}^\infty \frac{a_k}{\sum_{j=N+1}^\infty a_j} \mu_k)
$$
\begin{equation}
\geq (\sum_{i=1}^N a_i) \sum_{k=1}^N \frac{a_k}{\sum_{j=1}^N a_j} \ldim_E(\mu_k)
= \sum_{k=1}^N a_k \ldim_E(\mu_k).
\end{equation}
Since $N \in \N$ was arbitrary then
\begin{equation}
\ldim_E(\sum_{k=1}^\infty a_k \mu_k) \geq \sum_{k=1}^\infty a_k \ldim_E(\mu_k).
\end{equation}

We prove second inequality. It is well known that if $\nu \in \M1$ then
\begin{equation}
\udim_E(\nu) \leq \udim_B(X).
\end{equation}
As $\udim_B(X) < \infty$, for every $\e > 0$ we find $N \in \N$ such that 
\begin{equation}
\sum_{k=N+1}^\infty a_k \leq \frac{\e}{\udim_B(X)}.
\end{equation}
Thus by Theorem \ref{najwazniejszy}, we get:
\begin{equation}
\udim_E(\sum_{k=1}^\infty a_k \mu_k)
\leq (\sum_{i=1}^N a_i)\udim_E(\sum_{k=1}^N \frac{a_k}{\sum_{j=1}^N a_j} \mu_k) \end{equation}
$$
+ (\sum_{i=N+1}^\infty a_i)\udim_E(\sum_{k=N+1}^\infty \frac{a_k}{\sum_{j=N+1}^\infty a_j} \mu_k)
$$
\begin{equation}
\leq \sum_{k=1}^N a_k \udim_E(\mu_k) + \e 
\leq \sum_{k=1}^\infty a_k \udim_E(\mu_k) +\e.
\end{equation}
\end{IEEEproof}

Given a point $x \in X$, let $\d_x$ be an atomic measure at $x$, i.e.
\begin{equation}
\d_x(A):=\left\{
   \begin{array}{lcr}
		1 \mbox{, if }x \in A,\\ 
		0 \mbox{, if }x \notin A 
	\end{array}
\right.
\mbox{ for every } A \in \Sigma.
\end{equation}
Clearly, $\dim_E(\d_{x}) = 0$ for every $x \in X$. Making use of Theorem \ref{countable} we obtain the following corollary:
\begin{corollary} \label{corRen}
We assume that $\udim_B(X) < \infty$. Let $(x_k)_{k=1}^\infty \subset X$ and let $(a_k)_{k=1}^\infty \subset [0,1]$ be sequence such that $\sum\limits_{k=1}^\infty a_k = 1$. Then $\dim_E(\sum\limits_{k=1}^\infty a_k \d_{x_k}) = 0$.
\end{corollary}

 
\subsection{Improved Version of Young Theorem}

Finding the R\'enyi entropy dimension of a given measure is quite hard task in practice. It is much easier to calculate its local dimension. 

The local upper dimension of $\mu \in \M1$ at point $x \in X$, is defined by
\begin{equation}
\uD_\mu(x):=\limsup_{\d \to 0}\frac{\log\mu(B(x,\d))}{\log \d}.
\end{equation}

Fan \cite{Fa} obtained an estimation of upper entropy dimension of Borel probability measure on $\R^N$ by the supremum of local upper dimension, which can be seen as a version of Young Theorem \cite{Yo}:\medskip
\begin{flushleft}
\textbf{Consequence of Young Theorem} (see \cite[Theorem 1.3.]{Fa}) \textit{For a Borel probability measure $\mu$ on $\R^N$, we have}
\begin{equation}
\udim_E(\mu) \leq \esssup \uD_{\mu}(x).
\end{equation}
\medskip
\end{flushleft} 

We show that this estimation can be improved:
\begin{theorem} \label{twLok}
For a Borel probability measure $\mu$ on $\R^N$, we have
\begin{equation}
\udim_E(\mu) \leq \int_{\R^N}\uD_{\mu}(x) d\mu(x).
\end{equation}
\end{theorem}

\begin{IEEEproof}
Let us first observe that $\uD_\mu(x)$ is a measurable function, as the mapping $x \rightarrow \mu(B(x,\d))$ is measurable. 

Since for $\mu$-almost all $x \in \R^N$: $\uD_{\mu}(x) \leq N$ then we divide the segment $[0,N]$ into $n \in \N$ parts and denote sets
\begin{equation}
A_k^n:=\{x:\uD_\mu(x) \in (\frac{k-1}{n-1} N,\frac{k}{n-1} N]\}
\end{equation}
for $n \in \N$ and $k \in \{0,\ldots,n-1\}$. Let us define probability measures 
\begin{equation}
\mu_i^n:=\left\{
   \begin{array}{ll}
		\frac{1}{\mu(A_i^n)} \mu_{|A_i^n} & \mbox{, if }\mu(A_i^n) > 0,\\ 
		0 & \mbox{, if }  \mu(A_i^n) = 0
	\end{array}
\right.
\end{equation}
for $n \in \N$ and $i \in \{0,\ldots,n-1\}$. Since $A_i^n \subset X$ then
\begin{equation} \label{szacLok}
\uD_{\mu_i^n}(x) \leq \uD_\mu(x) \leq \frac{i}{n-1}N
\end{equation}
for $\mu$-almost all the points $x \in A_i^n$. Making use of Consequence of Young Theorem and (\ref{szacLok}), we have
\begin{equation} \label{lokal}
\udim_E(\mu_i^n) \leq \esssup \uD_{\mu_i^n}(x) \leq \frac{i}{n-1}N.
\end{equation}
By the definition of $\mu_i^n$, we represent measure $\mu$ as a convex combination of $\mu_i^n$, i.e. 
\begin{equation}
\mu = \sum_{i=0}^{n-1} \mu(A_i^n) \mu_i^n
\end{equation}
for each $n \in \N$.
Applying Theorem \ref{najwazniejszy} and (\ref{lokal}), we get
\begin{equation}
\udim_E(\mu) = \udim_E(\sum_{i=0}^{n-1} \mu(A_i^n) \mu_i^n) 
\end{equation}
\begin{equation}
\leq \sum_{i=0}^{n-1} \mu(A_i^n) \udim_E(\mu_i^n) 
\leq \sum_{i=0}^{n-1} \mu(A_i^n) \frac{i}{n-1} N.
\end{equation}
Finally taking limits as $n \to \infty$, we obtain
\begin{equation}
\udim_E(\mu) \leq \int_{\R^N} \uD_\mu(x) d\mu(x).
\end{equation}
\end{IEEEproof}

We were unable to verify whether a similar estimation holds for the lower entropy dimension, i.e. if $\int_{\R^N}\lD_{\mu}(x) d\mu(x) \leq \ldim_E(\mu)$.

\section{Conclusion}

Our paper investigates the problem of joint lossy-coding of information from combined sources. The main result gives the estimation of the entropy of mixture of sources by the combination of their entropies. The proof is based on the new equivalent definition of the entropy, which allows to obtain a convex combination of partitions contrary to the classical definition. We also present a practical and easy to implement algorithm of constructing joint coding alphabet for above problem. As a corollary we generalise some results concerning the R\'enyi entropy dimension.


%

\appendices
\section{Hardy-Polya-Littlewood Theorem}
We generalise the classical Hardy-Littlewood-Polya Theorem \cite[Theorem 1.5.4.]{Ni} for infinite sequences.

\medskip

\noindent{\bf Hardy-Littlewood-Polya Theorem.}
{\em
Let $a>0$ and let $\varphi:[0,a] \to \R_+$, $\varphi(0)=0$ be a continuous concave function.
Let $(x_i)_{i \in I}, (y_i)_{i \in I} \subset [0,a]$ be given sequences where
either $I=\N$ or $I=\{1,\ldots,N\}$ for a certain $N \in \N$.
We assume that
\begin{equation}
\sum_{i=1}^n x_i \leq \sum_{i=1}^n y_n \for n \in I
\end{equation}
and
\begin{equation}
\sum_{i \in I}x_i=\sum_{i \in I}y_i.
\end{equation}
If $(x_i)_{i \in I}$ is nonincreasing sequence then
\begin{equation} \label{cu}
\sum_{i \in I} \varphi(x_i) \geq \sum_{i \in I} \varphi(y_j).
\end{equation}
}

\begin{IEEEproof}
The classical Hardy-Littlewood-Polya Theorem \cite[Theorem 1.5.4]{Ni} covers exactly the finite sequence case, that is when $I=\{1,\ldots,N\}$ for a certain $N \in \N$. We will show that the case when $I=\N$ follows from the case when $I$ is finite.

To prove \eqref{cu} it is sufficient to show that
for every $n \in \N$ there exist $k_n \in \N$ such that
\begin{equation}
\sum_{i=1}^{k_n} \varphi(x_i) \geq \sum_{i=1}^{n} \varphi(y_i),
\end{equation}
since all sequences under considerations are nonnegative.
Let $n \in \N$ be arbitrary and let $k_n>n$ be chosen so that
\begin{equation}
r_{n+1}:=\sum_{i=1}^{k_n} x_i-\sum_{i=1}^{n} y_i \geq 0.
\end{equation}
Such a choice is possible since $(x_i)_{i \in I}$ and $(y_i)_{i \in I}$ are nonnegative sequences which
have equal sum.

Consider two finite sequences of equal length $k_n$:
$$
\tilde x=(x_1,\ldots,x_{k_n}) \mbox{ and }\tilde y=(y_1,\ldots,y_n,r_{n+1},0,\ldots,0).
$$
Observe that the above sequences have equal sum and that $\tilde x$ is nonincreasing.
We show that for every $k \leq k_n$
\begin{equation}
\sum_{i=1}^k \tilde x_i \leq \sum_{i=1}^k \tilde y_i.
\end{equation}
If $k \leq n$, this follows from the assumptions made on sequences $(x_i)_{i \in I}$ and $(y_i)_{i \in I}$.
If $k>n$ then
\begin{equation}
\sum_{i=1}^k \tilde x_i \leq \sum_{i=1}^{k_n} \tilde x_i= \sum_{i=1}^{k_n} x_i=
\sum_{i=1}^ny_i+r_{n+1}=\sum_{i=1}^{k} \tilde y_i.
\end{equation}
Since $(x_i)_{i \in I}$ is a nonincreasing we can apply to sequences $\tilde x$, $\tilde y$ and function $\varphi$ the finite sequence version of the classical Hardy-Littlewood-Polya and obtain that
\begin{equation}
\sum_{i=1}^{k_n} \varphi(x_i)=\sum_{i=1}^{k_n} \varphi(\tilde x_i)
\geq \sum_{i=1}^{k_n} \varphi(\tilde y_i)
\end{equation}
\begin{equation}
=\sum_{i=1}^{n} \varphi(y_i)+\varphi(r_{n+1})+
(k_n-(n+1))\varphi(0) 
\end{equation}
\begin{equation}
\geq \sum_{i=1}^{n} \varphi(y_i).
\end{equation}
\end{IEEEproof}

\ifCLASSOPTIONcaptionsoff
  \newpage
\fi




%
\bibliographystyle{IEEEtran}
\bibliography{IEEEfull,entropy}

%

\begin{IEEEbiographynophoto}{Marek \'Smieja}
received a master degree from mathematics at the Jagiellonian University, Krakow, Poland, in 2009, where he is currently working towards the Ph.D. degree at the Institute of Computer Science.
\end{IEEEbiographynophoto}

\begin{IEEEbiographynophoto}{Jacek Tabor}
received a master degree from mathematics at the Jagiellonian University, Krakow, Poland, in 1997. During the time period 1997-1998 he was on Fulbright Scholarship at the SUNY at Buffalo. In 2000 he obtained his Ph.D. in mathematics at the Jagiellonian University. Currently holds a professor position at the Institute of Computer Science of the Jagiellonian University.
\end{IEEEbiographynophoto}






\end{document}